\documentstyle[prl,epsfig,aps,twocolumn,amsmath]{revtex} 
\newcommand{\upd}{{\mathrm d}}  
\begin{document} 
 
\draft
\twocolumn[\hsize\textwidth\columnwidth\hsize\csname@twocolumnfalse\endcsname 
 
\title{Real space application of the
mean-field description of spin glass dynamics}

\author{Alain Barrat$^1$ and Ludovic Berthier$^2$
}  
  
\address{   
$^1$Laboratoire de Physique Th{\'e}orique  
\cite{umr}, B{\^a}timent 210, Universit{\'e} 
de Paris-Sud, 91405 Orsay Cedex, France \\ 
$^2$Laboratoire de Physique, ENS-Lyon and CNRS,
 F-69007 Lyon, France  and \\
D\'epartement de Physique des Mat\'eriaux, Universit\'e Claude Bernard
and CNRS, F-69622
Villeurbanne, France \\  } 
 
\date{\today} 
 
\maketitle 
\begin{abstract} 
The out of equilibrium dynamics of finite dimensional
spin glasses is considered from a point of view going beyond the standard 
`mean-field theory' versus `droplet picture' debate of the last decades.
The main predictions of both theories concerning the spin glass 
dynamics are discussed. 
It is shown, in particular, that predictions originating from mean-field ideas 
concerning the violations of the fluctuation-dissipation theorem 
apply quantitatively, provided one properly takes into account
the role of a spin glass coherence length, which plays a central role
in the droplet picture. 
Dynamics in a uniform magnetic field is 
also briefly discussed.
\end{abstract} 

\pacs{PACS numbers: 05.70.Ln, 75.10.Nr, 75.40.Mg \hspace*{4.4cm}
LPENSL-TH-01/2001}


\twocolumn\vskip.5pc]\narrowtext

Spin glasses have played a major and inspiring 
role in the rapidly developing field of the dynamics of glassy 
systems~\cite{young}.
In particular, important developments have been achieved in the study
of aging phenomena~\cite{review_aging}, 
which are encountered in several microscopic systems.
In this context, two very different dynamic descriptions
have emerged. The `mean-field theory' consists in the exact solution of 
the dynamics of fully-connected (or equivalently infinite-dimensional) 
spin glass models~\cite{review_aging}. 
The `droplet picture' is more phenomenological but directly 
addresses the problem of a real space (as opposed to the configurational
space) description~\cite{drop}.

It is a recurrent theme in the field to interpret
numerical or experimental data as validating one description 
at the expense of the other~\cite{enzo}.
Technical tools of the dynamic mean-field theory
have been argued to be necessary for a complete understanding of
the aging of finite dimensional spin glasses~\cite{review_simu}.
In this Letter, we show that their range of potential validity is 
in fact wider than previously thought (and can apply even when no
replica symmetry breaking is present), provided the theory 
is supplemented with the idea of a growing length scale, which we call
the `coherence length'.
A growing equilibration length, the `domain size', is at the heart of the 
description of aging phenomena by the droplet picture.
Although we cannot give a rigorous proof of their identity, we argue that 
these two length scales have the same physical content. 
We propose a construction for the coherence length and show
numerically that it is proportional to the domain size.
Last, we argue that the use of dynamical concepts coming
from mean-field ideas does not necessarily imply the existence 
of a static replica symmetry broken phase. 

The mean-field description of spin glass dynamics 
stems from the asymptotic solution of the dynamical equations for models 
which are statically solved by the Parisi replica symmetry breaking 
scheme~\cite{review_aging}. 
The behavior of the system is encoded in the autocorrelation function 
$C(t,t_w)$ and the conjugated response function $R(t,t_w)$.
It is shown that the decay of the correlation involves a complex
pattern of time scales organized in a hierarchical way. This 
`dynamic ultrametricity' is a direct counterpart of the static 
Parisi solution for these models~\cite{jpa}. However, this feature is absent 
from all known experimental and numerical data in three
dimensions~\cite{young,BBK2} which
show instead that the slow decay of the correlation
(or the thermoremanent magnetization $M(t,t_w)$ in experiments) is a 
one-time-scale process, $C(t,t_w) \simeq  {\cal C} (t/t_w)$
(or $M(t,t_w) \simeq  {\cal M} (t/t_w)$),
for times $t \gg t_w$.

Non-trivial predictions are also made concerning the relation between 
$R$ and $C$ which satisfy at equilibrium
the fluctuation-dissipation theorem (FDT), 
$T R(t,t_w)= \partial_{t_w} C(t,t_w)$.  
A generalization of the FDT is obtained by introducing
the function $X(t,t_w)$ through~\cite{jpa}
\begin{equation}
X(t,t_w) \equiv T R(t,t_w) \left( \frac{\partial C(t,t_w)}{\partial t_w}
\right)^{-1},
\end{equation}
with $X(t,t_w)=1$ at equilibrium.
In mean-field models, it can be shown that $X(t,t_w)$ becomes at long times 
a single argument function, allowing the definition
of the `fluctuation-dissipation ratio' (FDR) through~\cite{jpa}:
\begin{equation}
x(q) \equiv \lim_{t,t_w \to \infty} X(t,t_w) {\Big |}_{C(t,t_w)=q}\, .
\label{fdr}
\end{equation}
Moreover, this purely {\it out of equilibrium} quantity is related to the
spin glass order parameter $P(q)$,
which measures the {\it equilibrium} distribution 
function of overlaps~\cite{jpa,silvio2}, 
\begin{equation}
x(q) = \int_0^q \upd q' \, P(q').
\label{silvio}
\end{equation}

It has been further argued that Eqs.~(\ref{fdr})-(\ref{silvio}) are true
for finite dimensional glassy systems~\cite{silvio2}.
The existence of a FDR of the form (\ref{fdr}) has been 
numerically investigated in finite dimensional models~\cite{3dXx} 
through the study of the physically accessible quantities  
$C(t,t_w)$ and $\chi(t,t_w) \equiv \int_{t_w}^t \upd t'
R(t,t')$. 
Eq.~(\ref{fdr}) is then graphically checked by representing 
the variations of $\chi$ as a function of $C$ parameterized by the time
difference $t-t_w$, since Eq.~(\ref{fdr}) implies at large times
\begin{equation}
\chi(t,t_w) = \frac{1}{T} \int_{C(t,t_w)}^1 \upd q \, x(q), 
\label{parametric}
\end{equation}
i.e. the obtained $\chi(C)$ relation is independent of the time~\cite{jpa}.  
At equilibrium, $x=1$ and thus $\chi=(1-C)/T$.
Numerically~\cite{3dXx}, a $\chi(C)$ parametric curve 
is obtained for a large $t_w$ 
and compared to $S(C,L) \equiv \int_C^1 \upd q \int_0^q \upd q'  P(q',L)$ 
where $P(q,L)$ is the Parisi function computed in a system 
of linear size $L$ as large as possible.
The coincidence of these curves was used to argue
that both quantities were close to their limit, $t_w,L \to \infty$,
and to deduce informations on the nature of
the low-temperature phase~\cite{3dXx}.

A somewhat different picture has been put forward in Ref.~\cite{behose} and
checked in the case of the 2D XY model: a real space view of the aging 
behavior as an equilibration process taking place on a
growing `coherence length' 
$\ell(t_w)$ leads to a generalization of 
relations (\ref{silvio})-(\ref{parametric})  
at finite times and sizes:
\begin{equation}
T \chi(t,t_w) =  S(C(t,t_w),\ell(t_w)).
\label{conjecture}
\end{equation}
Eq.~(\ref{conjecture}) states that the {\it off-equilibrium properties}
of the infinite aging system at {\it finite-time} $t_w$ are connected to 
the {\it equilibrium properties} of a system of {\it finite-size} 
$\ell(t_w)$~\cite{behose2}, independently of the $t_w,L \to \infty$ limits.
In other words, the system is quasi-equilibrated up to a length scale
$\ell(t_w)$. For the 2D XY model, the coherence length is thus 
proportional to the dynamic correlation length $\ell(t_w) 
\approx \xi(t_w)$~\cite{behose,behose2}.

Inspired by dynamical mean-field theory, Equation~(\ref{conjecture}) 
{\it postulates} the existence of a growing coherence length $\ell(t_w)$
which represents the spatial extent to which the system is
equilibrated at time $t_w$.
Interestingly, this is the basic assumption of the droplet picture
that the spin glass dynamics is governed by 
large-scale excitations, whose size $\xi$ 
increases with time $t_w$ during the aging as 
$\xi(t_w) \sim (\ln t_w)^{1/\psi}$~\cite{drop}.
Thus, we expect $\ell(t_w) \sim \xi(t_w)$.

Numerically, $\xi(t_w)$ can be extracted from a four-point correlation 
function~\cite{3dXx,huse,heiko,hajime}. Its relevance for the scaling 
properties of various physical quantities has been
demonstrated in Refs.~\cite{hajime}.
Experimentally, $\xi(t_w)$ was
extracted through the scaling behavior of the Zeeman energy~\cite{zeeman}.
All these 3D investigations indicate a power law growth, 
$\xi(t_w) \sim t_w^{\alpha(T)}$, with $\alpha(T) \simeq 
0.2 \, T / T_g$. 

We now investigate numerically the validity of Eq.~(\ref{conjecture})
and then discuss its important consequences for the theoretical description 
of spin glasses.
Our aim is to demonstrate its wide range of applicability, independently
of the existence of replica symmetry breaking, in the context of which it
was first proposed~\cite{jpa,silvio2}. For this purpose, we consider
the low (but finite) temperature dynamics of the {\it 
two-dimensional} Edwards-Anderson model,
for which the spin glass transition is at $T_c=0$~\cite{heiko2}.
At low temperature, the relaxation time becomes so large
that the aging dynamics is very similar to the 3D case, including 
the power law growth  $\xi(t_w) \simeq t_w^{0.2 T}$~\cite{heiko}. 
The model is defined by
$H =  \sum_{\langle i,j \rangle} J_{ij} s_i s_j$,
where $s_i \, (i = 1,\cdots,N)$ are $N=L \times L$ Ising spins located  
on the sites of a square lattice of linear size $L$. 
The sum $\langle i,j \rangle$ runs over pairs of nearest neighbors.
The $J_{ij}$ are random Gaussian variables of mean 0 and variance 1.

\begin{figure}
\begin{center}
\psfig{figure=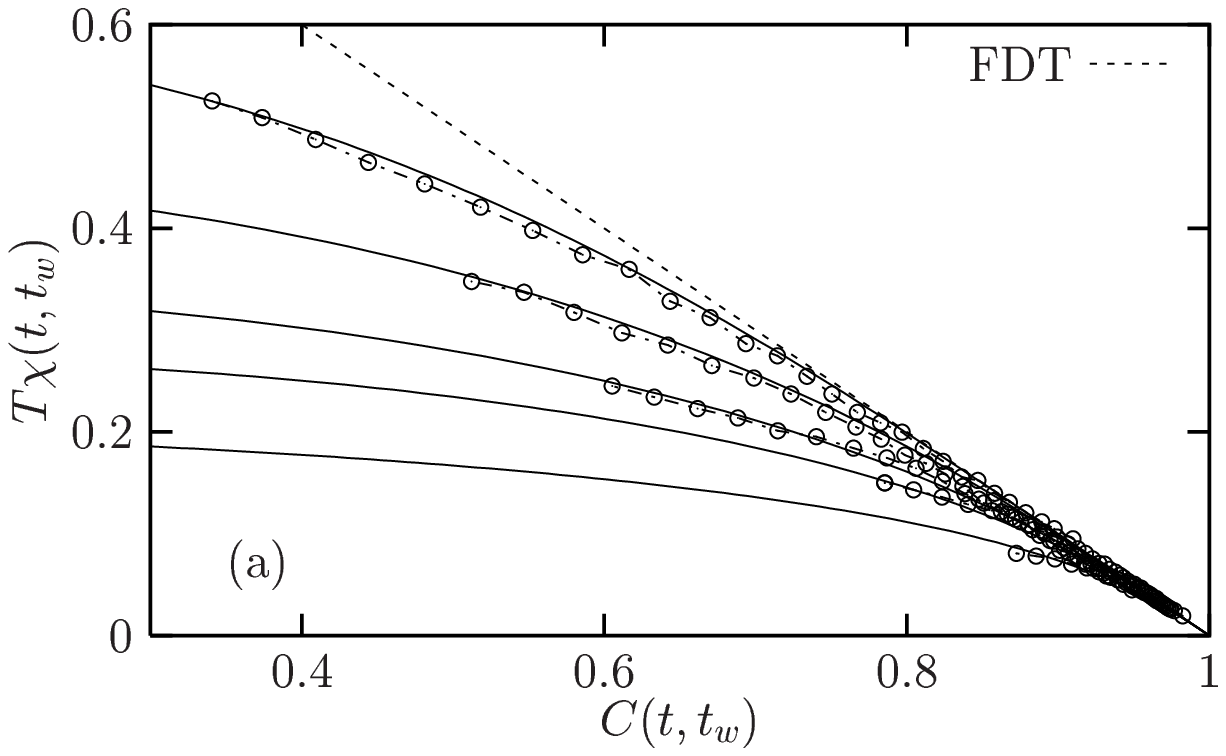,width=8.cm,height=5.5cm}
\psfig{figure=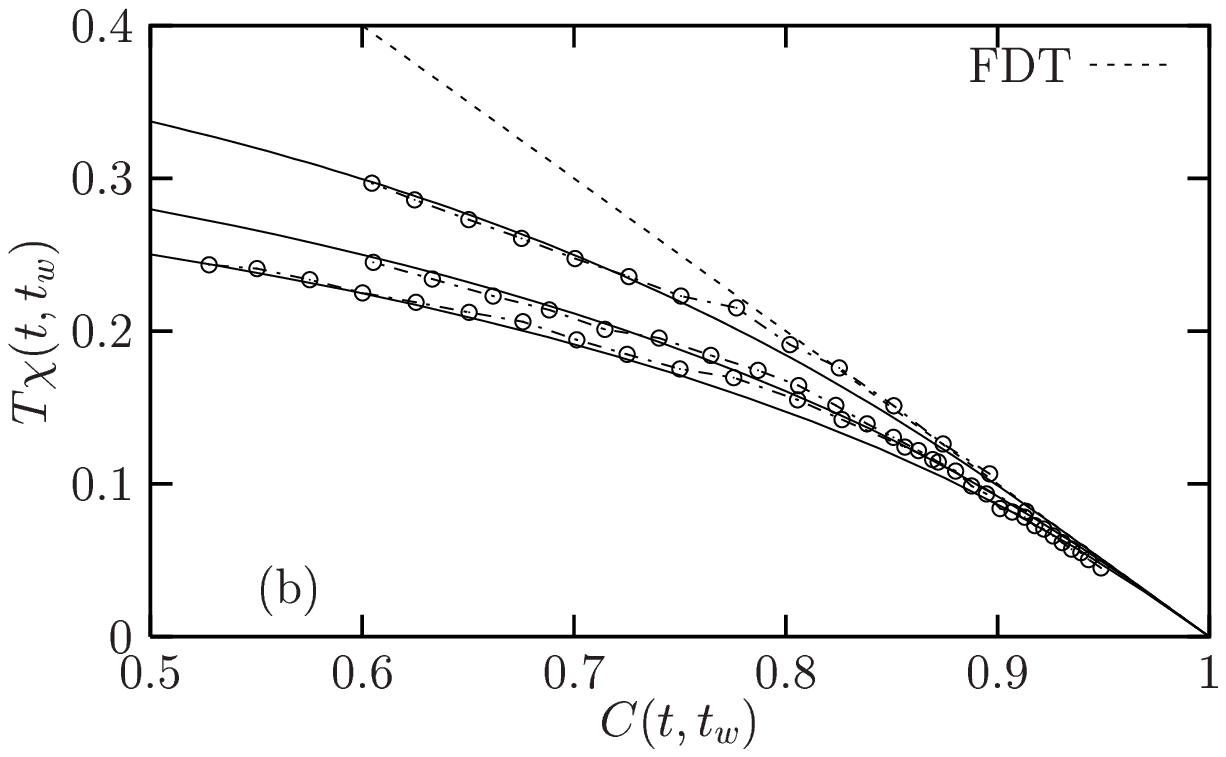,width=8.cm,height=5.5cm}
\caption{Susceptibility - Correlation parametric curves obtained in the aging
regime (circles) and from the static $S(C,L)$ (full lines).
The dashed line is the equilibrium FDT.
(a) Constant waiting time, $t_w=10^4$,
$T=0.6$, $0.5$, $0.4$, $0.3$, $0.2$ (from top to bottom).
(b) Constant temperature, $T=0.4$, and
$t_w=10^2$, $3.10^3$, $3.10^4$ (from bottom to top). Equilibrium data:
$L=8$, $10$, $18$ (from bottom to top).}
\label{conj}
\end{center}
\end{figure} 
\vspace*{-0.5cm}

To compute the autocorrelation function
$C(t,t_w)= N^{-1}  
\sum_i \overline{\langle s_i(t) s_i(t_w)\rangle}$
($\langle \, \cdots \, \rangle$ indicate an average over initial conditions
and $\overline{ \vphantom{()}\, \cdots \,}$ 
over realizations of the disorder),
and the susceptibility $\chi(t,t_w)$, 
a very large system, $L=400$, is quenched at the initial time 
$t_w=0$ from a disordered state to $T\in [0.2,1.0]$.
The susceptibility $\chi(t,t_w)$ 
is measured after applying a random magnetic field $h_i$
(taken from a Gaussian distribution
of mean 0 and variance ${h_0}^2$) between times
$t_w$ and $t$ in each site. 
In the linear response regime (we used $0.02 \leq h_0 \leq 0.05$),
one gets $\chi(t,t_w) = {h_0}^{-2} N^{-1}  
\sum_i \langle \overline{h_i s_i(t)} \rangle$.
The equilibrium $P(q,L)$ is computed independently by equilibrating (using 
parallel tempering~\cite{par_temp}) 
samples of sizes $L \in [6,24]$ and
temperatures $T\in[0.2,1.2]$.  
By definition, $P(q,L)$ is the disorder-averaged histogram of the overlap
$q = N^{-1} \sum_i s_i^a s_i^b$ between two equilibrated
copies $(a,b)$ of the system.
For the sizes and temperatures investigated,
$P(q,L)$ has its common `non-trivial' structure~\cite{review_simu}, with 
a peak around a $L$-dependent value of the `Edwards-Anderson parameter',
and a tail extending towards $q \simeq 0$ values, although 
we know that $\lim_{L\to\infty} P(q,L) = \delta(q)$ at all temperatures
$T>0$. 

We are now in position to compare the $\chi(C)$ 
curves obtained in the aging situation, with $S(C,L)$.
Our results are summarized in Fig.~\ref{conj}. 
We show first in Fig.~\ref{conj}-a the parametric curves for 
the same large waiting time, $t_w=10^4$, and different temperatures. 
The dynamic curves are qualitatively similar 
to the 3D case [to our knowledge,
no such data are available in 2D].
At short times ($1-C \ll 1$), the curves follow the equilibrium 
FDT relation, which they 
smoothly leave at longer times.

The validity of Eq.~(\ref{conjecture}) is demonstrated in two steps.
Fig.~1-a shows first that the dynamic curves follow, within our numerical
precision, the curves of $S(C,L)$ obtained by the double 
integration of $P(q,L)$, for a given value of $L$.
Note that the coincidence of these two independently computed functions
on their whole support is a very strong requirement which 
unambiguously defines  the coherence length, $\ell(t_w) = L$.
The time evolution of $\ell(t_w)$ is then followed for $T=0.4$
in Fig.~1-b which shows that dynamic curves for increasing
$t_w$ coincide with static curves for increasing sizes.
For all temperatures investigated, we find a growth law 
for $\ell(t_w)$ consistent with the 2D growth laws reported in 
Ref.~\cite{heiko} for $\xi(t_w)$, showing that
$\ell(t_w) \propto \xi(t_w)$.
Physically, the coincidence between dynamic and static data means that
at time $t_w$, the system is locally equilibrated up to a coherence length
$\ell(t_w)$.
The precise link between $\ell(t_w)$ and $\xi(t_w)$ is however
a tricky point since
the static $P(q,\ell(t_w))$ is sensitive to the boundary
conditions~\cite{review_simu}.
We use here (as is usual~\cite{3dXx})
periodic boundary conditions which leads to the 
estimate $\ell(t_w) \simeq 2 \xi(t_w)$. 
The equality $\ell(t_w)=\xi(t_w)$ would probably be obtained 
computing $P(q,L)$ in a box of size $L$ inside a much larger system,
as proposed in Ref.~\cite{newstein}.
This is of course much more computationally demanding.

At very low temperature ($T \lesssim 0.3$), we find that the agreement
between statics and dynamics is not as good, with slightly
different shapes for static
and dynamic data. This is probably because even at very large times,
the coherence length is very small ($\ell \simeq 1 - 3$),
so that the regime where Eq.~(\ref{conjecture}) becomes valid
is not reached. However, a similar trend has been 
observed in Ref.~\cite{behose}, 
attributed to the role of topological defects. No data are available 
yet in 3D or 4D spin glasses at very low temperature~\cite{3dXx}, 
and this point should be checked. 

The dynamic data of Fig.~\ref{conj}-a obtained
at different temperatures can be collapsed
using the scaling variables $A\equiv \chi T^{1-\Phi}$ and
$B\equiv (1-C)T^{-\phi}$, with $\Phi$ a free parameter
chosen to get the best collapse of the curves in Fig.~\ref{fig:collapse}.
For $1-C \ll 1$ (short times), the FDT 
implies $B = A$, while we obtain at large times the power law
$B \sim A^{1-1/\Phi} \simeq A^{0.565}$, i.e. $\Phi \simeq 2.3$.
This scaling behavior has been proposed in Ref.~\cite{3dXx} as a dynamic 
analog for the so-called Parisi-Toulouse approximation used in a 
mean-field static context. 
The relation $B \sim A^{0.41}$ has been numerically found in 3D and 4D
spin glasses, and compared to an expected mean-field
behavior $B \sim A^{0.5}$~\cite{3dXx}.
That the scaling works also in 2D explicitly shows that 
it is not necessarily connected to an underlying replica symmetry breaking,
and weakens the evidence presented in other numerical works.

\vspace*{-0.2cm}
\begin{figure}
\begin{center}
\psfig{figure=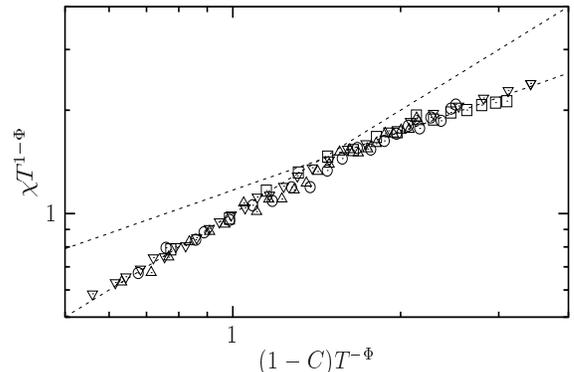,width=7.5cm,height=5.cm}
\caption{Scaling behavior of the dynamic curves of Fig.~\ref{conj}-a
with $\Phi\simeq 2.3$, guessed from the Parisi-Toulouse approximation.
The dashed lines are the relations
$B=A$ and $B \sim A^{0.565}$.}
\label{fig:collapse}
\end{center}
\end{figure}
\vspace*{-1cm}
\begin{figure}
\begin{center}
\psfig{figure=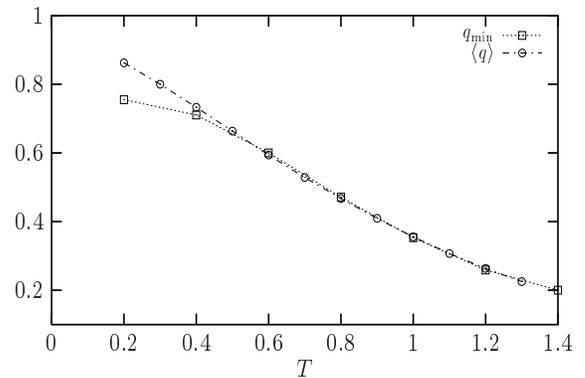,width=7.5cm,height=5.cm}
\caption{Temperature dependence
of the overlaps $\langle q \rangle$ and $q_{min}$, in a 
uniform magnetic field $h=0.4$.}
\label{fig:at}
\end{center}
\end{figure}

\vspace*{-0.6cm}
We discuss now different important implications of the
finite-time / finite-size connection described by Eq.~(\ref{conjecture}).
From a pragmatic point of view, first, this relation implies that 
numerical studies of large aging systems or small equilibrated systems
potentially contain the same informations. 
In our opinion, this fact has been largely underestimated in the spin glass
literature, which often tries to overcome the difficulty of
obtaining thermalized samples of large sizes by using large times
in dynamical simulations. We exemplify this point by discussing the behavior 
of spin glasses in a uniform magnetic field.
Various static tests of the existence of a spin glass phase
give inconclusive results ($P(q,L)$ in a field
differs from its mean-field shape, Binder cumulants do not 
cross in a field~\cite{review_simu}), while 
the dynamic behavior in a field 
has been claimed to clearly demonstrate the existence of a 
replica symmetry broken phase, in 3D and 4D~\cite{ea3_4dfield}.
This intriguing fact led us to perform in 2D the simulations of  
Refs.~\cite{ea3_4dfield}.
We have computed different values of the overlap, $q_{min}$
and $\langle q \rangle \equiv \int \upd q' \, P(q',L)q'$ 
which, according to the mean-field
shape of $P(q,L=\infty)$ in a field, 
should be different~\cite{ea3_4dfield};   
$q_{min}$ is obtained dynamically as the 
infinite time extrapolation of the overlap between two independently 
aging copies of the system~\cite{ea3_4dfield}.
Fig.~\ref{fig:at} shows that, as in 3D and 4D, the two values differ
at low enough $T$. 
Taken at face value, this result would lead to the 
erroneous conclusion that a replica symmetry breaking transition 
occurs at $T_c(h=0.4) \simeq 0.5$\cite{kinzelbinderyoung}. 
This shows that dynamic studies, using the same computer resources,
are in fact probing the same length scales as equilibrium ones 
and therefore have to be taken with the same care. 
Much larger times (in dynamics) and 
sizes (in statics) should be used to  show that there is no transition 
in a field in 2D.

The validity of Eq.~(\ref{conjecture}) implies that the same 
excitations of length $\ell(t_w)$ are governing the dynamics, i.e. the decay
of the correlation as $C(t,t_w) \simeq {\cal C}(\ell(t)/\ell(t_w))$,
and are contributing to the low-$q$ part of $P(q,L)$, i.e.
to the non-FDT part of the parametric curves in Fig.~\ref{conj}.
We emphasize that Eq.~(\ref{conjecture}) would not be satisfied 
at finite time during the coarsening in a pure ferromagnet (below
$T_c$) where
the domain walls governing the aging are absent 
in the static situation with usual boundary conditions.
(See the related discussion of the role of vortices in Ref.~\cite{behose}.) 
Hence, we prefered the name `coherence length' 
to `domain size', since these two names refer to qualitatively different
physical situations.
Our results give no indication on the structure of the relevant
excitations in spin glasses, but underlines the relevance 
(and the need) of finite-$T$ equilibrium studies of large-scale excitations  
for the description of aging. It is not established, in particular, 
that the description~\cite{drop} in terms of compact droplets is 
correct~\cite{martin}.

From a fundamental point of view, our results give 
support to the picture
of aging in spin glasses as the successive equilibration of 
excitations of increasing length scales, recently 
put forward to interpret temperature-cycling experiments~\cite{JP}.
We speculate that the absence of $\ell(t_w)$ in mean-field theory and 
hence of the resulting multi-length-scale dynamics,
is ``compensated'' by the dynamic ultrametricity~\cite{jpa}.
This results in the inability of the theory of correctly 
predicting the simple $t/t_w$-scaling of dynamic functions.
Note that the reported growth laws are inconsistent
with the logarithmic law predicted by scaling arguments~\cite{drop,JP},
which would anyway lead to the unobserved 
$\log t /\log t_w$-scaling.
A thorough investigation of this law seems necessary, and
current experiments and simulations~\cite{JP2} should clarify
this point. 

Note finally that the extremely slow growth 
of the coherence length implies that the length scales 
involved in the dynamics are relatively small,  
{\it even at experimental times}. 
The provocative 
idea that the thermodynamic limit might be of no 
{\it practical} importance directly follows.
This also implies that Eq.~(\ref{conjecture}) should apply
in experiments, leading to results quantitatively
similar to Fig.~\ref{conj}.

We thank J.-P. Bouchaud, C. De  Dominicis, P. Holdsworth, 
J. Kurchan, M. M\'ezard, M. Moore, F. Ricci-Tersenghi,
F. Zuliani for discussions. 
This work was supported by the PSMN at ENS-Lyon.


\begin{thebibliography}{99}
\bibitem{behose2}
For 2D XY model, the interpretation is clear~\cite{behose}: 
at $t_w=0$, critical fluctuations of length scales between
the lattice and the system sizes are present in an out of equilibrium
state. After a time $t_w$, fluctuations of length
scales smaller than $\ell (t_w) \simeq {t_w}^{1/z}$ have equilibrated, while
the others still have the memory of their non-equilibrium initial conditions.
\end{thebibliography}
\end{document}